\documentclass[prb,
twocolumn,
superscriptaddress,showpacs,amsmath,amssymb]{revtex4}

\begin{document}

\title{Double Hawking temperature in de Sitter Universe and cosmological constant problem}

\author{G.E.~Volovik}
\affiliation{Low Temperature Laboratory, Aalto University,  P.O. Box 15100, FI-00076 Aalto, Finland}
\affiliation{Landau Institute for Theoretical Physics, acad. Semyonov av., 1a, 142432,
Chernogolovka, Russia}

\date{\today}

\begin{abstract}
{
As distinct  from the black hole physics, the de Sitter thermodynamics is not determined by the cosmological horizon, the effective temperature  differs from the Hawking temperature. In particular,  the atom in the de Sitter universe experiences thermal activation corresponding to the local temperature, which is twice larger than the Hawking temperature,  $T_{\rm loc}=2T_{\rm Hawking}$. The same double Hawking temperature describes the decay of massive scalar field in the de Sitter universe. The reason, why the local temperature is exactly twice the Hawking temperature, follows from the geometry of the de Sitter spacetime. The weakening of the role of the cosmological horizon in de Sitter universe is confirmed by considering Hawking radiation.
We discuss the difference between the radiation of particles in the de Sitter spacetime and the Schwinger pair creation in the electric field. We use the stationary Painleve-Gullstrand metric for the de Sitter spacetime, where the particles are created by Hawking radiation from the cosmological horizon, and  time independent gauge for the electric field. In these stationary frames the Hamiltonians and the energy spectra of massive particles look rather similar. However, the final results are essentially different.  
In case of Schwinger pair production the number density of the created pairs grows with time, while in the de Sitter vacuum 
the number density of the created pairs is finite. The latter suggests that Hawking radiation from the cosmological horizon does not lead to instability of the de Sitter vacuum. The other mechanisms of instability are required for the dynamical solution of the cosmological constant problem.   We consider the possible role of the local temperature  $T_{\rm loc}=2T_{\rm H}$ in the decay of the de Sitter space-time due to the energy exchange between the vacuum energy and relativistic matter with this temperature.
}
\end{abstract}
\pacs{
}

\maketitle 
 
 \section{Introduction}

The issue of the stability of the de-Sitter vacuum is still an unsolved problem. This problem is related to the also unsolved problem of cosmological constant and dark energy. The particular solution of the cosmological constant problem is provided by the $q$-theory.\cite{KlinkhamerVolovik2008}  Following Hawking,\cite{Hawking1984} the $q$-theory describes the dynamics and thermodynamics of the  dark energy in terms of the 4-form field. In thermodynamics, the $q$-theory demonstrates, that in the equilibrium Minkowski vacuum there is the natural cancellation of the vacuum energy without any fine tuning. 

However, the problem remains in dynamics of vacuum energy, where the result of the relaxation of  the initial state with the large dark energy depends on the stability or instability of the de Sitter expansion. If the de Sitter expansion is stable and does not dissipate, then  both  the Minkowski and de Sitter vacua serve as the attractors in the vacuum dynamics.  To exclude the de Sitter attractor some mechanism of the decay of the de Sitter vacuum is necessary. In other words, while the $q$-theory solves cosmological constant problem for the equilibrium vacuum, the decay of the de Sitter vacuum is the necessary condition for the dynamical solution of the cosmological constant problem within this theory.\cite{KlinkhamerVolovik2017} 

When studying the stability of the de-Sitter expansion, it is necessary to take into account the important differences between the properties of the ideal de Sitter expansion and that of the perturbed de Sitter, when it is deformed by matter fields; by the expansion history; or by other factors, which violate the de Sitter symmetry. 

This symmetry is violated, for example, by an atom living in the background of de Sitter vacuum. The interaction of the atom with the de Sitter vacuum leads to ionization of the atom in the absence of electric field, the process which is not possible in the Minkowski vacuum due to energy conservation.  The process of ionization is very similar to the ionization of atoms  by a weak external electric field in the flat spacetime.\cite{Volovik2009b} This example demonstrates the analogy between  the de Sitter vacuum and the Minkowski vacuum with electric field. Both processes can be studied using the semiclassical description of quantum tunneling. In the de Sitter case, the tunneling exponent corresponds to the thermal ionization with the effective temperature, which  is twice the traditional Hawking temperature describing radiation from the event horizon:
\begin{equation}
 T_{\rm loc}=2T_{\rm H} \,.
\label{localT}
\end{equation}

Eq.(\ref{localT}) is valid in the regime, when  the relevant classical trajectories of electron in the ionization process are well within the event horizon. This means that effective temperature is the local temperature, and thus this process of ionization has no relation to the cosmological horizon.  The same local temperature describes the process of  the splitting of the composite particle with mass $M$ into two components with $M_1 +M_2>M$, which is also not allowed in the Minkowski vacuum.\cite{Bros2008,Bros2010,Volovik2009b} The decay rate of massive field with the same Boltzmann factor $\exp{(-M/T_{\rm loc})}$ has been obtained in Ref. \cite{Jatkar2012}, see also Ref.\cite{Maldacena2015}. The physical origin of the factor  2 in Eq.(\ref{localT}) for the local temperature $T_{\rm loc}$, which describes the thermal processes occurring within the cosmological horizon, will be discussed in Sec. \ref{TlocalSec}.

That the Hawking temperature and the cosmological horizon  are not relevant for the de Sitter thermodynamics is supported by the consideration of the Hawking radiation. For the perfect de Sitter vacuum, there is the analogy between the Hawking radiation and the Schwinger creation of charged particle in a uniform electric, see e.g. Refs. \cite{Mottola2017,Polyakov2018}. In Sec.\ref{dSvsSchwinger} we will show that there is the principle difference between these two processes. We compare these two processes using the approach of semiclassical tunneling.\cite{JacobsonVolovik1998,Padmanabhan1999,Volovik1999a,ParikhWilczek,Parikh2002,Padmanabhan2002,Volovik2009a,Volovik2009b}
As distinct from Ref.\cite{Mottola2017}, we consider the time independent gauge for the electric field and the stationary Painleve-Gullstrand metric\cite{Painleve,Gullstrand}  for the de Sitter spacetime, where the particles are created by Hawking radiation from the cosmological horizon. In these stationary frames the two systems look rather similar. However, the final results are essentially different.  
In case of Schwinger pair production the number density of the created pairs grows with time, while in the de Sitter vacuum 
the number density of the created pairs is finite. The latter suggests that Hawking radiation from the cosmological horizon does not lead to instability of the de Sitter vacuum, and supports the weakening of the role of the cosmological horizon in the de Sitter thermodynamics.  

The other mechanisms of instability are required for the dynamical solution of the cosmological constant problem. 
In Sec.\ref{dSdecay} we consider the possible role of the local temperature  $T_{\rm loc}=2T_{\rm H}$ in the decay of the de Sitter space-time due to the energy exchange between the vacuum energy and relativistic matter with this temperature.

 \section{de Sitter entropy and Hawking temperature versus local temperature}
 \label{TlocalSec}

Here we consider how the perturbed de Sitter expansion leads to effective thermodynamics with the local temperature $T_{\rm loc}$
in Eq.(\ref{localT}), which is the double Hawking temperature.

 \subsection{Double Hawking temperature}
 \label{DiubleSec}
 
The interaction of the atom with the de Sitter vacuum leads to ionization of the atom in the absence of electric field, the process which is not possible in the Minkowski vacuum due to energy conservation.  The tunneling exponent corresponds to the thermal ionization, and the effective temperature (\ref{localT}) is twice the Hawking temperature for radiation from the de Sitter cosmological horizon,\cite{Volovik2009b} 

Eq.(\ref{localT}) is valid in the regime, when  the relevant classical trajectories of electron in the semiclassical  ionization process are well within the event horizon. This means that effective temperature is the local temperature, and thus this process of ionization has no relation to the cosmological horizon and to the Hawking radiation.  

The same local temperature describes the process of  the splitting of the composite particle with mass $M$ into two components with $M_1 +M_2>M$, which is also not allowed in the Minkowski vacuum.\cite{Bros2008,Bros2010,Volovik2009b} The probablity of this process  (for $M\gg T_{\rm H}$):
\begin{eqnarray}
\Gamma(M \rightarrow M_1+M_2) \sim \exp{\left(-\frac{\pi (M_1+M_2-M)}{H}\right)} =
\nonumber
\\=  \exp{\left(-\frac{ M_1+M_2-M}{ T_{\rm loc}}\right)}\,.
\label{mto2}
\end{eqnarray}

If only single scalar field with mass $M$ is present, the process of the splitting of a particle to two particles becomes the process of duplication of particles, i.e. the particle transforms to  two particles of the same bosonic field, $M \rightarrow 2M$. The equation  (19) in Ref.\cite{Volovik2009b} gives 
for $M_1=M_2=M$ the probability of this process
\begin{equation}
\Gamma(1 \rightarrow 2) \sim \exp{\left(-\frac{\pi (2M-M)}{H}\right)}= \exp{\left(-\frac{ M}{T_{\rm loc}} \right)} \,.
\label{mto2m}
\end{equation}
This coincides with the decay rate of massive field  $\Gamma(1\rightarrow 0) \sim  \exp{\left(-\frac{M}{T_{\rm loc}} \right)}$ obtained in Ref. \cite{Jatkar2012}. This demonstrates that creation of new particles  is compensated  by their decay,
 $n\Gamma(1\rightarrow 2)=n\Gamma(1\rightarrow 0)$, where $n$ is particle density. This detailed balance, however, does not determine  the distribution function $n$ in equilibrium.

 The equilibrium particle density can be found from the consideration of Gibbons-Hawking radiation from de Sitter cosmological  horizon.
In the Hawking radiation, two particles are created: one is inside horizon and the other is outside. For the local observer at the cosmological horizon the Gibbons-Hawking radiation   is observed as the coherent creation of two particles with mass $M$. Thermal excitation of two particles corresponds the environment with temperature, which is double Hawking temperature, $T_{\rm loc}=2T_{\rm H}$, i.e. 
$\Gamma(0 \rightarrow 2) = \exp{\left(-\frac{M}{T_{\rm H}} \right)}=  \exp{\left(-\frac{2M}{T_{\rm loc}} \right)}$. Since the de Sitter state is symmetric, all the points in its space are equivalent, and thus at each point the Hawking radiation is viewed as the creation of pairs from the vacuum.
The detailed balance $\Gamma(0 \rightarrow 2) =n\Gamma(1\rightarrow 2)= n \Gamma(1 \rightarrow 0)$ gives $n \sim  \exp{\left(-\frac{M}{T_{\rm loc}}\right)}= \exp{\left(-\frac{M}{2T_{\rm H}} \right)}$.

The factor $\exp{\left(-\frac{M}{2T_{\rm H}} \right)}$ is discussed in detail in Ref. \cite{Maldacena2015} devoted to the so-called Cosmological Collider (inflation viewed as a particle accelerator).
It is stressed that this factor differs from conventional thermal factor  $\exp{\left(-\frac{M}{T_{\rm H}}\right)}$, which characterizes the creation of a pair of massive particles.

Let us note here that, though the Hawking radiation itself is not relevant for the de Sitter universe, its consideration allows us to determine the local temperature in terms of the Hawking temperature, $T_{\rm loc}=2T_{\rm H}$.

 \subsection{Local thermodynamics}
 \label{ThermodynamicSec}

Another argument, which supports the uniqueness of the perfect de Sitter expansion and its insensitivity to the cosmological horizon,  is unusual value of the entropy of the dS state. 
The thermodynamics of the dS vacuum suggests that its entropy is zero.\cite{Pollock1989}
If the thermodynamics is described in terms of the Hawking temperature of radiation from the cosmological horizon, $T_{\rm H}= H/2\pi$, then the positive contribution 
$S_+= \frac{1}{4}A/G$ from the region beyond the horizon is compensated by the negative contribution from the interior, $S_-= -\frac{1}{4}A/G$. 

This cancellation can be also interpreted using the result of Ref. \cite{Volovik2020}, where it was found that the white hole has negative  entropy. The entropy of the white hole with the same mass as the black hole is $S_{\rm WH}=-S_{\rm BH}$. The reason for that is the opposite sign of the area of their horizons, $A_{\rm WH}=-A_{\rm BH}=-4\pi r_{\rm hor}^2$. The same can be applied for the interior and exterior of the cosmological horizon. They have the same boundary (horizon) but viewed from different sides, and thus $A_-=-A_+$.

However, the compete cancellation of contributions from the opposite sides of horizon suggests that the presence of the horizon is not important both for the entropy and for Hawking radiation discussed in Sec. \ref{dSvsSchwinger}, i.e. consideration of the de Sitter entropy in terms of the event horizon and Hawking radiation from this horizon is rather artificial. In Sec.\ref{DiubleSec} we considered another approach, which rely on the geometry of the de Sitter expansion rather than on the cosmological horizon.  The matter fields in de Sitter universe experience the local temperature, which is different from the Hawking temperature related to horizon. This temperature, which is twice the Hawking temperature, $T_{\rm loc}=H/\pi=2T_{\rm H}$, is measured by the local observer.\cite{Volovik2009b} In particular, the quantum tunneling process of decay of the composite particle in the de Sitter vacuum well within the horizon looks as thermal process with the effective temperature $T_{\rm loc}$.\cite{Bros2008} 

Let us consider how this local temperature $T_{\rm loc}$ enters statistical physics of the de Sitter universe.
According to Ref. \cite{Podolskiy2018}  the probability of fluctuation of the potential of the scalar field $\Phi$ in the de Sitter spacetime is
\begin{equation}
P \propto \exp{ \left(- \frac {8\pi^2}{3H^4} \delta V(\Phi) \right)} =\exp{ \left(- \frac {F(\Phi)}{T} \right)}\,,
\label{fluctuations}
\end{equation}
where $F$ is the free energy of the matter field. 

There are two ways to interpret this probability. The traditional interpretation is in terms of the Hawking temperature: \cite{Podolskiy2018}
\begin{equation}
 T=T_{\rm H}=\frac{H}{2\pi}\,\,\,,\,  F(\Phi)=\frac{4\pi}{3H^3} \delta V(\Phi) \,.
\label{interpretation1}
\end{equation}
Here the free energy is determined in the region within the cosmological horizon with volume  $\frac{4\pi}{3H^3}$, i.e.  the volume of a given Hubble patch. 

On the other hand, Eq.(\ref{fluctuations}) describes the fluctuations of the matter fields on the background of the de-Sitter vacuum. That is why they should be discussed in terms of the physical temperature, i.e. the local temperature experienced by the matter fields, $T=T_{\rm loc}$. 
Then  from Eq.(\ref{fluctuations}) one obtains 
\begin{equation}
 T=T_{\rm loc}=\frac{H}{\pi}\,\,\,,\,  F(\Phi)=\frac{8\pi}{3H^3} \delta V(\Phi) \,.
\label{interpretation2}
\end{equation}
Now the free energy is determined in the whole volume of the expanding patch.
The static coordinates with Hubble volume $\frac{4\pi}{3H^3}$ inside the horizon cover only half of the expanding Poincare patch. The extension to the whole expanding Poincare patch requires doubling  of the effective volume, and this is the origin of the volume $\frac{8\pi}{3H^3}$  in Eq.(\ref{interpretation2}). This interpretation is in terms of the local physics and does not depend on the existence of the event horizon, which artificially separates the patches inside and outside the horizon. 

Note that the tunneling process of the particle decay, which is characterized by the effective temperature $T=T_{\rm loc}$, is not possible in the pure de Sitter spacetime, since the particle is simply absent in the pure de Sitter. Both the decaying particle  and the observer violate the symmetry of the pure de Sitter spacetime, which makes the process possible. 

So the thermodynamics of fluctuations on the background of the de Sitter is fully determined by the local temperature, which is not related to the cosmological horizon. The reason, why this physical temperature is twice larger than the Hawking temperature, follows from the geometry  of the de Sitter space.

 \section{de Sitter vs Schwinger in semiclassical approximation}
 \label{dSvsSchwinger}
  
  The other argument, that the Hawking radiation from cosmological horizon with $T=T_{\rm H}$ probably is not important for the de Sitter universe, comes from consideration of particle production in the pure de Sitter vacuum.
  Let us compare the particle production in the perfect de Sitter vacuum and in the constant electric field in flat space using the semiclassical approximation. This approximation is valid when the Hubble parameter  and the electric field are small compared to the mass of particles, $H\ll M$ and ${\cal E}\ll M^2$ correspondingly. In this semiclassical approximation there is no difference between fermions and bosons, except for the integer number in prefactor (number of species, spin, polarization, etc.). In both cases we use the description in terms of the stationary metric and fields.

   \subsection{Semiclassical energy spectrum of particle in electric field and in de Sitter spacetime}

We use the Painleve-Gullstrand metric known in condensed matter as acoustic metric:\cite{Unruh1976}
\begin{equation}
ds^2= - dt^2(1-{\bf v}^2) - 2dt\, d{\bf r}\cdot {\bf v} + d{\bf r}^2 \,,
\label{PGmetric}
\end{equation}
where ${\bf v}({\bf r})$ is the velocity of the free-falling observer, who crosses the horizon, and $c=1$.  In the de Sitter case the velocity 
${\bf v}({\bf r})=H{\bf r}$, where $H$ is Hubble parameter, and the cosmological horizon is at $r_{\rm hor}=1/H$.

In the Painleve-Gullstrand metric, the energy spectrum  of particles  is Doppler shifted: 
$E \rightarrow E +  {\bf p}\cdot {\bf v}({\bf r})$. In the semiclassical approximation, the  spectrum of particle with mass $M$ in the de Sitter spacetime is:
\begin{equation}
E({\bf p},{\bf r})= \pm \sqrt{M^2 +p^2} + H {\bf p}\cdot {\bf r}  \,.
\label{dSSpectrum}
\end{equation}
This can be compared with the spectrum of a charged particle in a constant electric field:
\begin{equation}
E({\bf p},{\bf r})= \pm \sqrt{M^2 +p^2} + {\bf {\cal E}}\cdot {\bf r} \,,
\label{dSSchwinger}
\end{equation}
where electric charge $q=1$ is assumed.

The main difference between Eq.(\ref{dSSpectrum}) and Eq.(\ref{dSSchwinger}) is that in the de Sitter case the electric field 
${\bf {\cal E}}$ is substituted by $H {\bf p}$.

Let us now neglect the curvature of cosmological horizon at $Hr=1$. Then introducing the coordinate  $z$ across the cosmological horizon:
\begin{equation}
E({\bf p},z)= \pm \sqrt{M^2 +p_\perp^2+p_z^2} + H p_z z \,.
\label{dSSpectrumZ}
\end{equation}
Introducing correspondingly  the coordinate  $z$ along the electric field, one obtains for Schwinger case:
\begin{equation}
E({\bf p},z)= \pm \sqrt{M^2 +p_\perp^2+p_z^2} + {\cal E} z \,.
\label{dSSchwingerZ}
\end{equation}

  \subsection{Tunneling approximation}
 
 In the semiclassical approximation, the probability  of the particle creation is given by the tunneling exponent:
 \begin{equation}
W=\sum_{\bf p}w_{\bf p}=\sum_{\bf p}\exp{\left(- 2{\rm Im}\int dz p_z(z) \right)}\,.
\label{Probability}
\end{equation}
where the tunneling trajectories along $z$ are given by equation $E({\bf p},z)=E$.
In case of Schwinger pair production the tunneling exponent  depends only on the transverse momentum ${\bf p}_\perp$:
\begin{equation}
w_{\bf p}^{\rm Schwinger}=\exp{\left(- \frac{\pi \tilde M^2}{\cal E} \right)}\,\, , \,\, \tilde M^2=M^2 +p_\perp^2 \,,
\label{SchwingerExp}
\end{equation}
while in the de Sitter case the tunneling exponent depends also on the longitudinal momentum:
\begin{equation}
w_{\bf p}^{\rm dS}=\exp{\left(- \frac{2\pi E}{H}\right)} \,\, , \,\, E^2=M^2 +p_\perp^2 + p_{z0}^2 \,,
\label{dSexp}
\end{equation}
The semiclassical approximation is valid for correspondingly $M \gg H$ and $M^2 \gg {\cal E}$.

  \subsection{de Sitter vs Schwinger}
   \label{dSnodecay}
  
Let us first consider the Schwinger pair production.
Integration over transverse momenta gives
\begin{equation}
\int \frac{d^2 p_\perp}{(2\pi)^2}\exp{\left(- \frac{\pi \tilde M^2}{\cal E} \right)}= 
 \frac{\cal E}{(2\pi)^2}  \exp{\left(- \frac{\pi  M^2}{\cal E} \right)} \,.
\label{SchwingerExpIntegrate}
\end{equation}
Integral over $dp_z /2\pi$ diverges because the exponent does not depend on $p_z$. Due to the motion equation  $dp_z = {\cal E}dt$, one obtains the known Schwinger pair creation per unit volume per unit time (the integer factor for polarization and spin of particles  is ignored):
\begin{equation}
\Gamma^{\rm Schwinger} = \frac{dW^{\rm Schwinger}}{dt} = 
 \frac{{\cal E}^2}{(2\pi)^3}  \exp{\left(- \frac{\pi  M^2}{\cal E} \right)} \,.
\label{SchwingePairCreation}
\end{equation}

On the other hand, for the de Sitter case the integral over $p_z$ is finite, and one obtains the total probability of the particle production per unit volume:
\begin{eqnarray}
W^{\rm dS}=\int \frac{d^3p}{(2\pi)^3} \exp{\left(- \frac{2\pi E}{H}\right)}= 
\label{dSexpIntegrate1}
\\
=  \frac{\sqrt{2}}{\pi^3}(MH)^{3/2}  \exp{\left(- \frac{2\pi M}{H}\right)} \,.
\label{dSexpIntegrate2}
\end{eqnarray}
That is why in the limit of infinite time the creation of particles per unit time is zero, $\Gamma^{\rm dS}=dW^{\rm dS}/dt=0$.
This is in contrast to $\Gamma^{\rm dS}\sim M^3H \exp{\left(- \frac{2\pi M}{H}\right)}$ obtained in Ref.\cite{Mottola2017} and suggests that 
the Hawking radiation does not lead to relaxation of the de Sitter vacuum. In this respect the non-perturbed de Sitter expansion represents the unique time-dependent gravitational field, which is not radiating even in the presence of the cosmological horizon.

The combined effects, when pair creation takes place in curved spacetime in the presence electric field, have been considered earlier, see e.g. Refs. \cite{Buchbinder1982,Vilenkin2014,Pedraza2015}

 \section{Double Hawking temperature and de Sitter decay}
 \label{dSdecay}
 
 If the Hawking radiation does not lead to the decay of the de Sitter expansion, the other mechanisms of instability are required for the dynamical solution of the cosmological constant problem. In Sec.\ref{TlocalSec}  we considered the unconventional thermodynamics of the de Sitter vacuum, and found that it is characterized by the effective temperature, which is twice larger than the Hawking temperature, $T_{\rm loc}=2T_{\rm Hawking}$. This follows from the specific geometry of the de Sitter expansion, and is not related to the existence of the cosmological horizon. In other words, the de Sitter thermodynamics demonstrates that it can be described without consideration of the cosmological horizon and Hawking radiation. 
Though the Hawking radiation itself is not relevant for the de Sitter universe, its consideration however allows us to determine the local temperature in terms of the Hawking temperature, $T_{\rm loc}=2T_{\rm H}$.

 The  quantum tunneling process, which leads to the decay of the composite particle in the de Sitter vacuum, occurs fully inside the cosmological horizon and is fully determined by the local temperature, $T_{\rm loc}$. The free energy of the fluctuations of the matter fields also corresponds to the local temperature, $T_{\rm loc}$. It is possible that the presence of matter and its nonzero local temperature  $T_{\rm loc}=2T_{\rm H}$ may cause the relaxation of the vacuum energy. Let us consider this scenario.
 
 We consider the final stage, when the cold dark matter practically disappears and there is only the dark energy.
 The local nonzero temperature acting on matter field  $T_{\rm loc}=H/\pi$ suggests that it generates  the thermal relativistic gas with  $T=T_{\rm loc}$ and thus with the energy density $\rho_M =bH^4$, where  the dimensionless parameter $b$ depends on the number of the massless relativistic fields. 
 
 The relativistic matter fields in the de Sitter spacetime obey the conservation law 
 \begin{equation}
\partial_t \rho_M= - 4H \rho_M \,,
\label{MatterConservation}
\end{equation}
according to which the matter density decreases with time. The constant value of $\rho_M = bH^4$, which is caused by the local temperature $T_{\rm loc}=H/\pi$, is only possible if there is the energy flux from the dark energy to the relativistic matter, which must compensate its energy losses. With the energy transfer the dynamical Friedmann equations become:
\begin{eqnarray}
\partial_t \rho_M= - 4H (\rho_M - b H^4)\,,
\label{MatterNonConservation}
\\
\partial_t \rho_V= - 4b H^5 \,.
\label{VacuumNonConservation}
\end{eqnarray}

Since the vacuum energy density is $\rho_V \propto H^2/G$, one obtains from Eq.(\ref{VacuumNonConservation}):
 \begin{eqnarray}
H \sim E_{\rm Pl} \left( \frac{t_{\rm Pl}}{t}\right)^{1/3}  \,,
\label{DecayLawH}
\\
\rho_V \sim E_{\rm Pl}^4  \left( \frac{t_{\rm Pl}}{t}\right)^{2/3} \,,
\label{DecayLawV}
\\
\rho_M =bH^4 \sim  E_{\rm Pl}^4  \left( \frac{t_{\rm Pl}}{t}\right)^{4/3}  \,,
\label{DecayLawM}
\end{eqnarray}
 where the Planck time $t_{\rm Pl}^2=G$ and Planck energy $E_{\rm Pl}=1/t_{\rm Pl}$ are introduced.
 Here it is assumed that $t\gg t_{\rm Pl}$ and thus $\rho_M \ll \rho_V$. That is why, in the main approximation the dark energy is not disturbed, and $\partial_t H \ll H^2$.
 
The power law decay of $H$ in Eq.(\ref{DecayLawH}) is similar to that in Eq.(109) in Ref. \cite{Markkanen2018} 
with initial value $H_0 \sim E_{\rm Pl}$. Such decay may take place at the very early time, when the cold matter is absent. The relativistic matter decays much slower than due to the exponential expansion.
The dark energy  decays to zero, as follows from thermodynamics. However, the decay rate is too slow to explain the small value  $\rho_V \sim E_{\rm Pl}^4  \left( \frac{t_{\rm Pl}}{t_{\rm present}}\right)^2$ at the present time $t_{\rm present}$.
 
This regime can take place also at the very late time, when the cold matter already disappears. Taking into account the value of the dark energy at present time, one obtains  that this regime can emerge only at the highly unrealistic time (see also Ref.\cite{Dvali2017}):
  \begin{equation}
 t > t_{\rm present} \left( \frac{t_{\rm present}}{t_{\rm Pl}}\right)^2 \sim 10^{120} t_{\rm present}
 \,.
\label{time}
\end{equation}

 In both cases such decay cannot explain the dynamical origin of the present small value of the cosmological constant, though such small value follows from the thermodynamics.\cite{KlinkhamerVolovik2008}
 
In the $q$-theory,\cite{KlinkhamerVolovik2008} at the early time the vacuum relaxation is mainly determined by the oscillations. It is either exponentially fast, or experiences the power law decay, $\rho_V \sim E_{\rm Pl}^4  \left( \frac{t_{\rm Pl}}{t}\right)^2$. The latter gives the correct estimate of the present value of the cosmological constant, if the Universe approaches Minkowski vacuum. If instead the Universe approaches the de Sitter state, the final stage of the decay will be governed by Eq.(\ref{DecayLawV}). 
 
  \section{Conclusion}
  
We discussed the  possibility of the decay of  the de Sitter state, and considered two possible sources of the decay: the 
Hawking radiation from the cosmological horizon and the local temperature, $T_{\rm loc}=2T_{\rm H}$, which in particular characterizes the  decay of the composite particles in the de Sitter spacetime.

To study the possible effect of the cosmological horizon and Hawking radiation, we compared  the Hawking radiation in the de Sitter spacetime with the Schwinger pair production in flat space. Using the stationary spacetime for the de Sitter expansion and the semiclassical approximation for quantum tunneling, we obtained  that the Hawking radiation across the cosmological horizon is described in the same way as the Schwinger pair production in the uniform electric field. However, the final results are different. For the  Schwinger pair production the semiclassical approximations reproduces  the known  result for the intensity of the pair production $\Gamma^{\rm Schwinger}$. For the de Sitter case the production of pairs per unit volume, $W^{\rm dS}$, is finite and thus the intensity (the production per unit volume per unit time) is $\Gamma^{\rm dS} =dW^{\rm dS}/dt=0$. 
 
 This does not mean that the de Sitter vacuum is stable: this only means that the Hawking radiation alone does not lead to instability, i.e the de Sitter vacuum is stable with respect to
the decay via the Hawking radiation. The Hawking radiation does not lead to the change of the vacuum energy density, which generates the de Sitter expansion. This means that even if the pair creation takes place, the de Sitter expansion immediately dilutes the produced particles, and thus there is no vacuum decay in de Sitter.  The possibility of the decay of the pure de Sitter vacuum due to Hawking radiation requires the further consideration.

There are many other mechanisms, not related to the Hawking radiation, which could lead to the decay of the de Sitter spacetime,\cite{Starobinskii1977,Starobinskii1979,Starobinskii1983,StarobinskyYokoyama1994,Polyakov2008,Polyakov2012,Akhmedov2014,Palti2019}
including the infrared instability, instability due to the dynamic effects of a certain type of quantum fields, instability towards spontaneous breaking of the symmetry of the de Sitter spacetime or the instability towards the first order phase transition in the vacuum, etc.
But in most cases either the de Sitter vacuum is not perfect, i.e. there are deviations from the exact de Sitter, and the de Sitter symmetry is lost, or the vacuum energy is fine-tuned, i.e. the cosmological constant problem is ignored. The de Sitter instability, which avoids fine tuning, but uses the special vector field in Dolgov scenario,\cite{Dolgov1997} is in Ref.  \cite{Emelyanov-Klinkhamer2012}.

 We considered the possible effect of the local temperature, $T_{\rm loc}=2T_{\rm H}$. If the matter fields in the de Sitter spacetime are characterized by this temperature, then there should be the energy exchange between the vacuum and relativistic matter given by Eqs. (\ref{MatterNonConservation}) and (\ref{VacuumNonConservation}). This leads to the decay of the vacuum energy to zero, but the rate of the decay is too slow to explain the present value of the cosmological constant.

The problem of the dynamical stability of the de Sitter vacuum is directly related to the cosmological constant problem. The $q$-theory\cite{KlinkhamerVolovik2008}   demonstrates the solution of the problem in thermodynamics: in the equilibrium Minkowski vacuum the cosmological constant is nullified due to thermodynamics. However, the relaxation of the de Sitter state to the equilibrium vacuum depends on the  stability of the de Sitter vacuum.  If the de Sitter attractor is not excluded in dynamics, then the only possibility to solve the dynamical cosmological problem within the $q$-theory is to assume that the Big Bang occurred in the part of the Universe, which is surrounded by the equilibrium environment.\cite{KlinkhamerSantillan2019}

{\bf Acknowledgements}. I thank A. Polyakov and A. Starobinsky for discussions.
This work has been supported by the European Research Council (ERC) under the European Union's Horizon 2020 research and innovation programme (Grant Agreement No. 694248).

%%%%%%%%%%%%%%%%%%%%%%%%%%%%%%%%%%%%%%%%%%
\end{document}